\documentstyle[prb,aps]{revtex}
\begin{document}
\draft
\title{Electronic Structure of Single- and Multiple-shell Carbon Fullerenes}

\author{Yeong-Lieh Lin and Franco Nori}
\address
{Department of Physics, The University of Michigan, Ann Arbor, MI 48109-1120}

\maketitle
\begin{abstract}
We study the electronic states of giant single-shell and the recently
discovered nested multi-shell carbon fullerenes within the
tight-binding approximation. We use two different approaches, one based
on iterations and the other on symmetry, to obtain the $\pi$-state
energy spectra of large fullerene cages: $C_{240}$, $C_{540}$,
$C_{960}$, $C_{1500}$, $C_{2160}$ and $C_{2940}$. Our iteration
technique reduces the dimensionality of the problem by more than one
order of magnitude (factors of $\sim 12$ and $20$), while the
symmetry-based approach reduces it by a factor of $10$. We also find
formulae for the highest occupied and lowest unoccupied molecular
orbital (HOMO and LUMO) energies of $C_{60{\cdot}n^{2}}$ fullerenes as
a function of $n$, demonstrating a tendency towards metallic regime for
increasing $n$. For multi-shell fullerenes, we analytically obtain the
eigenvalues of the intershell interaction.

\end{abstract}


\narrowtext

{\em Introduction.---}The discovery of a new simple
technique\cite{Krats} for the production in bulk quantities of
fullerenes has triggered intensive research on these new all-carbon
molecules and the search for other novel forms of carbon. Subsequently,
a new type of carbon structure composed of multilayered needle-like
tubes has been discovered by high-resolution transmission electron
microscopy\cite{Ijima}. Quite recently, giant nested shells of
onion-like fullerenes (also called hyperfullerenes\cite{Curl}) have
been synthesized by intense electron-beam irradiation\cite{Ugarte}.
Essentially they consist of a composition of concentric spherical
fullerene cages. The innermost cage is a $C_{60}$ molecule which is
encapsulated by giant fullerenes $C_{240}$, $C_{540}$, $\cdots$, one
following the other\cite{Ugarte,Kroto}. The interlayer spacing
coincides with that for bulk graphite ($3.34$ \AA). A recent
investigation\cite{Tomanek} on the stability of these brand-new diverse
forms of carbon shows that multi-sh Specifically, a very recent
letter\cite{Maiti} shows that a stability transition from single to
multilayer fullerenes occurs when the number of atoms exceeds $\sim
6000$. Moreover, the concentric carbon onion structure containing no
dangling bonds provides a challenge to graphite (comprising flat sheets
of carbon hexagons) for the most stable form of pure carbon.

The structures and energies for several carbon ``onions" have recently
been carried out by using realistic atomic potentials\cite{Maiti} and
molecular mechanics calculations\cite{Yoshida}. As yet there has been
little spectroscopic information available regarding the electronic
properties of the multi-shell fullerenes, besides total energy
calculations. The exact diagonalization of a local orbital matrix
scales as $N^{3}$ (for a $N-$dimensional matrix). The
orthonormalization step in plane-wave methods also scales as $N^{3}$.
These two examples illustrate typical bottlenecks encountered when
studying one of the central thrusts of condensed matter theory: to
compute the energetics of very large systems. Different suggestions on
how to minimize this formidable problem have recently attracted a lot
of attention (see, {\em e.g.}, Ref.~9). Here, we explore different
ways to reduce this difficulty by using two very dissimilar approaches:
one based on iterations and the other on symmetry considerations. {\em
The iteration method reduces the dimensionality of the problem, $N$, by
more than one order of magnitude} ({\em factors of $\sim$ either $12$
or $20$
depending on the chosen initial state}). For instance, it reduces $N$
by about a factor of $20$, from $2940$ to $154$, for the $C_{2940}$
fullerene molecule. Furthermore, it can provide some precise analytic
results for the eigenvalues. By using a different method based only on
symmetry arguments, we can reduce $N$ by a factor of $10$.  These two
approaches give the same results for the energy eigenvalues. However,
several distinct features of the first method that make it very
convenient include the following: it requires the diagonalization of a
fewer number of matrices, each one of them smaller in size, and every
one of them with all their elements real.

Here we focus on a simple model of the electronic states of the giant
single-shell fullerenes, and multiple-shell carbon ``onions" with the
number of shells up to seven. The strategy used in this work includes
two parts. First, we treat each $C_{60{\cdot}n^{2}}$ fullerene as a
$60{\cdot}n^{2}$-site nearest-neighbor tight-binding model. Each site
corresponds to one radially directed carbon $\pi$ orbital. Since the
$\pi$ orbitals provide the dominant contribution to the conducting
properties of the molecules, the $\sigma$ orbitals (being mainly
responsible for the elastic properties) are not taken into account. In
spite of the slight differences between bond lengths and structure
curvature in these molecules, we have assumed a uniform hopping
amplitude $t$ for all the fullerenes under consideration.
 This assertion is motivated by the fact that results in $C_{60}$
 changed insignificantly by considering the two different hopping
 integrals. These icosahedrally-symmetric $C_{60{\cdot}n^{2}}$
 molecules will be studied for $n=1,2,\cdots,7$ ({\em i.e.}, $C_{60}$,
 $C_{240}$, $C_{540}$, $C_{960}$, $C_{1500}$, $C_{2160}$ and
 $C_{2940}$).

In the last part of this work, we use an intermolecular tight-binding
model to account for the weak intershell interaction between the
fullerene layers and analytically obtain its eigenvalues. It is
important to stress that the simple tight-binding approximation yields
reasonable quantitative results for the energy levels of each
individual fullerene\cite{Mano,paco,Yang}, especially for those near
the Fermi level. Furthermore, the intermolecular tight-binding model
has also been applied to the electronic states of solid $C_{60}$.

{\em Single-shell giant fullerenes.---}In the description of the
electronic states of $C_{60{\cdot}n^{2}}$ molecules, we employ a
tight-binding Hamiltonian (H\"{u}ckel theory as, for instance, used for
fullerenes in Refs.~10-12) written as $H=- \sum_{<ij>} t\
c_{i}^{+}\,c_{j}$.  Admittedly, this is a simplified Hamiltonian: the
``Ising model" for the electronic structure. However, its understanding
is a convenient stepping stone to the study of more complex models.
Also, simple models are more accessible for analytical or
semi-analytical approaches. Much more realistic and complicated
tight-binding Hamiltonians can be found, for instance, in Ref.~13 and
papers cited therein. For carbon $C_{60}$, we have obtained closed-form
expressions of the eigenvalues and eigenfunctions for both $\pi$ and
$\sigma$ states as well as the Green function and local density of
states by using recursion and path-integral methods\cite{Lin}. Here we
focus on the giant fullerenes.

To obtain the energy spectrum, we use two entirely different
approaches, each one with its advantages. The first approach we use
here is our generalization of the recursion method\cite{Haydock}, which
we call the ``generalized block recursion method". Because of limited
space, here we present only a brief outline of this approach. We start
from two initial states, $|u_{0}>$ and $|v_{0}>$, formed as
\begin{equation}
|u_{0}>=\frac{1}{\sqrt{10}}\,\sum_{j=1}^{5}\ [|j>+ {\cal P}\,|j'>]
\end{equation}
and
\begin{equation}
|v_{0}>=\frac{1}{\sqrt{12}}\,\sum_{j=1}^{6}\ (-1)^{j+1}[|j>+ {\cal P}\,|j'>].
\end{equation}
In  $|u_{0}>$ ($|v_{0}>$), $|j>$ denotes the $\pi$ orbitals on a
pentagon (hexagon), and $|j'>$ denotes those on the opposite ({\em
i.e.}, $j$ and $j'$ are antipodes) pentagon (hexagon). Also ${\cal P}$
stands for parity, with the value $+1$ or $-1$. Notice that these two
pentagon (hexagon) rings satisfy the prerequisite that a  fivefold
(threefold) symmetry axis passes through their centers. Further states
are then iteratively generated by the general recurrence relation
\begin{equation}
H\,|f_{m}^{\alpha}>=a_{m}^{\alpha}\,|f_{m}^{\alpha}>+b_{m+1}^{\beta}\,|f_{m+1}^{\beta}>+b_{m}^{\gamma}\,|f_{m-1}^{\gamma}>,
\end{equation}
where $m$ designates states which are $m$ hops away from the initial
state. The superscript designates various suitable states. On the
left-hand side of Eq. (3), at most three terms are nonzero for each
application of $H$ on $|f_{m}^{\alpha}>$. Among them, $\beta$ and
$\gamma$ can each take at most two values among their several possible
ones. The process terminates after a finite number of iterations. The
representation of $H$ thus becomes a ``block-tridiagonal" form.
However, the dimensionality of each block is {\em not} necessarily the
same. The main operational difference between this approach and the
standard recursion method is that additional indices ($\alpha$, $\beta$
and $\gamma$) are needed to fully specify the generated states,
producing very simple values ({\em i.e.}, $\pm 1$, $\pm 2$ and
$\sqrt{2}$) for the coefficients $a$'s and $b$'s. Furthermore, the
standard recursion method is extremely cumbersome to apply for
$C_{60{\cdot}n^{2}}$ (for $n>1$), while very convenient for only
$n=1$.

The advantage of this approach lies in that it reduces the task of
diagonalizing the extremely large original Hamiltonian matrix for
$C_{60{\cdot}n^{2}}$ to the diagonalization of four much smaller
matrices and provides precise analytic results for some of the
eigenvalues ({\em e.g.}, $\pm 1$, $\pm 2$, $-3$). For instance, in
order to obtain the whole energy spectrum of $C_{240}$, we only need to
diagonalize two (due to ${\cal P}=\pm 1$) $14 \times 14$ (constructed
from $|u_{0}>$) and two $18 \times 18$ (constructed from $|v_{0}>$)
matrices instead of a $240 \times 240$ matrix. Thus, the dimensionality
has been reduced by (factors of $\sim 12$ and $20$) more than one order
of magnitude.
 In general, for $C_{60{\cdot}n^{2}}$ the dimensionality is reduced
 from $60n^{2}\ $ to $3n^{2}+n\ \ (5n^{2}-n)$ starting from the initial
 state $|u_{0}>\ (|v_{0}>)$. The relative size reduction is
 $(3n^{2}+n)/60n^{2} = 1/20 + 1/60n\ {\simeq}\ 1/20$$\ $
 $(\ (5n^{2}-n)/60n^{2} = 1/12 - 1/60n\ {\simeq}\ 1/12)$ for
 $|u_{0}>\ (|v_{0}>)$. Therefore, the dimensionality of the original
 giant fullerene problem can be easily reduced by more than one order
 of magnitude. The LUMO (HOMO) energy is solved from the matrix
 constructed from $|u_{0}>$ ($|v_{0}>$) with negative parity.

We also use another approach, which closely follows the method
presented in Ref.~12. This alternative and quite different method is
based on group theory analysis. For instance, for $C_{240}$, six $24
\times 24$ matrices (four of them with complex elements) are
diagonalized to obtain the energy spectrum. This method reduces the
dimensionality of the problem by a factor of $10$. For comparison
purposes, we also use a third different method for the smaller cages:
direct exact diagonalization. This method runs into limitations for
very large cages. Therefore, it is a desirable and useful goal to be
able to first reduce the dimensionality of the problem by more than one
order of magnitude, in order to study very large fullerene cages.

The energy levels from the first two methods for  various fullerenes
are shown in Fig.~1. In Table I, we summarize the following physically
important quantities for the giant fullerenes: energies of the HOMO,
LUMO and their differences ({\em i.e.}, the band gaps). As the size of
the fullerene grows, it can be seen that  the band gap becomes smaller.
These results demonstrate the tendency towards metallic regime for very
large fullerenes. The best logarithmic fit formulae of the HOMO and
LUMO energies (in units of $t$) for $C_{60{\cdot}n^{2}}$ molecules with
$n$ up to 7 are
\begin{equation}
E_{{\rm HOMO}}=-0.67402\cdot n^{-0.72956}
\end{equation}
and
\begin{equation}
E_{{\rm LUMO}}=0.14670\cdot n^{-1.3772}.
\end{equation}
By taking $t$ equal to the typical value of $2.5$ eV, these expressions
become $E_{{\rm HOMO}}=-1.68505\cdot n^{-0.72956}$\ eV and $E_{{\rm
LUMO}}=0.36675\cdot n^{-1.3772}$\ eV. Even though these fits are not
very accurate for small $n$, we believe that Eqs.~(4) and (5) provide
good approximations for larger values of $n$ ($n \geq 8$) where
calculations are very difficult. From them, the band gap can also be
readily inferred.

{\em Multi-shell carbon fullerenes.---}In this section, we study the
effects of the interaction between consecutive shells on the electronic
states of the fullerene shells. In the stackings of the onion-like
structures, the pentagons are aligned\cite{Yoshida} along the
icosahedral directions of the twelve vertices, and the hexagons are
stacked in a manner similar to the AB (or Bernal) structure of
graphite. Nevertheless, and as verified in Ref.~8, the interlayer
interaction is mainly between adjacent shells. Each single-shell
fullerene is now assumed to be in its ground (lowest energy) state and
here we focus on its HOMO.
 We assume an intermolecular hopping amplitude $V_{n}$ between the pairs
of fullerenes $C_{60{\cdot}n^{2}}$ and $C_{60{\cdot}(n+1)^{2}}$. A
tight-binding approximation, $H_{I}=- \sum_{n=1}^{N_{s}-1}
V_{n}\,(\,|\phi_{n}><\phi_{n+1}|+|\phi_{n+1}><\phi_{n}|\,),$ is then
used to model the intershell interaction, where $|\phi_{n}>$ denotes
the HOMO of the $C_{60{\cdot}n^{2}}$ molecule and $N_{s}$ is the total
number of shells. The $V_{n}$'s can be obtained, for example, through
an intermolecular resonance integrals calculation.
 At this moment, such calculations cannot yet be accurately done because
of the lack of knowledge about the exact atomic positions. However, the
$V_{n}$'s are estimated to be of the order of a few hundred meV. Notice
that while the multiplicity of the HOMO for each fullerene is five, the
nonzero matrix element only exits between those belonging to the same
representation. In other words, molecular states $|\phi_{n}>$ and
$|\phi_{n+1}>$ in  $H_{I}$ must be in the same representation. Thus,
for a $N_{s}$-shell carbon ``onion", a small $N_{s} \times N_{s}$
matrix (constructed from $H_{I}$), with the $m$-th row and $n$-th
column element $-V_{n-1}{\delta}_{m,n-1}-V_{n}{\delta}_{m,n+1}$, is
sufficient to fully investigate the  interaction between the $5N_{s}$
HOMOs. The eigenvalues for $H_{I}$ can then be analytically solved. The
results for various $N_{s}$ are presented in Table II.  This model with
different hopping amplitudes can be readily applied to the threefold
degenerate LUMOs.

In summary, we use two different approaches to study the $\pi$-state
energy spectra of fullerene cages up to $C_{2940}$. We also find
formulae for the HOMO and LUMO energies of fullerenes
$C_{60{\cdot}n^{2}}$ as a function of $n$. These approximations might
be useful since experimentally obtained carbon ``onions" can have $\sim
70$ shells (with $294,000$ atoms for the $70$th shell, these systems
are well beyond current computational capabilities). We obtain the gap
energy as a function of $n$ and show the tendency towards metallic
regime for very large fullerenes. For multi-shell carbon ``onions", we
analytically obtain the eigenvalues of the intershell  interaction. The
HOMO, LUMO, and gap energies, as well as the whole energy spectra, of
these molecules are relevant to the several important experimental
techniques which probe the spectroscopy of single molecules. For
instance, a scanning tunneling microscope ({\it e.g.}, Ref.~16) can be
used to probe the local spectroscopy of fullerenes.

The authors gratefully acknowledge stimulating and useful discussions
with S. Fahy. FN acknowledges partial support from a GE fellowship, the
NSF through grant DMR-90-01502, and SUN Microsystems.

\begin{figure}
\caption{The energy eigenvalues (horizontal axis) and their
corresponding degrees of degeneracy (vertical axis) for various
fullerenes. The energies are in units of carbon-carbon hopping integral
$t$. The respective fivefold degenerate HOMO and threefold degenerate
LUMO levels are indicated. For $C_{60{\cdot}n^{2}}$, there are
approximately $60{\cdot}n^{2}/4\ =\ 15{\cdot}n^{2}$ distinct energy
levels.}
\label{fig1}\end{figure}

\begin{figure}
\caption{The HOMO, LUMO and gap energies (in units of eV) versus $1/n$  for
various fullerenes $C_{60{\cdot}n^{2}}$.}
\label{fig2}\end{figure}

\newpage
\widetext

\begin{table}
\caption{$N_{L}=3n^{2}+n$\ \ ($N_{H}=5n^{2}-n$) is the dimensionality
of the matrix constructed from $|u_{0}>$\ \ ($|v_{0}>$), which provide
the LUMO (HOMO) energies. These are significantly smaller than the
dimensionality $N=60n^{2}$ of the original Hamiltonian. Energies of the
HOMO, LUMO and band gaps for $C_{60{\cdot}n^{2}}$ fullerenes, with
$n=1, 2, \cdots, 7$. In the HOMO and LUMO columns, the first values are
in units of $t$ and those in parentheses are the results obtained by
taking $t$ equal to the typical value of $2.5$ eV.}
\begin{tabular}{lcccccccccc}
\multicolumn{1}{c}{Molecule}& \multicolumn{1}{c}{$N_{L}$} &
\multicolumn{1}{c}{$N_{H}$} &\  &\multicolumn{2}{c}{HOMO}  & \   &
\multicolumn{2}{c}{LUMO} & \  &Gap (eV) \\ \hline
$C_{60}$ & $4$& $4$&\ & $-0.618033$ & ($-1.545082$) & \ & $0.138564$ &
($0.346410$) &
\ & $1.891492$ \\
$C_{240}$ & $14$& $18$&\ & $-0.436772$ & ($-1.091930$) & \ & $0.059657$ &
($0.149142$) &
\ & $1.241072$ \\
$C_{540}$ & $30$& $42$&\ & $-0.324099$ & ($-0.810247$) & \ & $0.033671$ &
($0.084177$) &
\ & $0.894424$ \\
$C_{960}$ & $52$& $76$&\ & $-0.255010$ & ($-0.637525$) & \ & $0.022169$ &
($0.055422$) &
\ & $0.692947$ \\
$C_{1500}$ & $80$& $120$&\ & $-0.209564$ & ($-0.523910$) & \ & $0.015971$ &
($0.039927$) &
\ & $0.563837$ \\
$C_{2160}$ & $114$& $174$&\ & $-0.177673$ & ($-0.444182$) & \ & $0.012192$ &
($0.030480$) &
\ & $0.474662$ \\
$C_{2940}$ & $154$& $238$&\ & $-0.154140$ & ($-0.385350$) & \ & $0.009690$ &
($0.024225$) &
\ & $0.409575$ \\
\end{tabular}
\label{table1}
\end{table}

\narrowtext
\begin{table}
\caption{The energy eigenvalues of $H_{I}$ for various $N_{s}$.}
\begin{tabular}{cc}
$N_{s}$ & Eigenvalues \\ \hline
$2$ & $\pm V_{1}$ \\
$3$ & $0$\ \ \ \ \ \ \ \ and\ \ \ \ \ \ \ \ $\pm \sqrt{V_{1}^{2}+V_{2}^{2}}$ \\
$4$ & $-{\cal A}\pm {\cal B}$\ \ \ \ \ \ \ \ and\ \ \ \ \ \ \ \ ${\cal A}\pm
{\cal B}$ \\
$5$ & $0$,\ \ \ \ \ \ \ \ $\pm \sqrt{{\cal S}-{\cal T}}$\ \ \ \ \ \ \ \ and\ \
\ \ \ \ \ \ $\pm \sqrt{{\cal S}+{\cal T}}$ \\
\end{tabular}
\label{table2}
\tablenotetext{Here ${\cal A}=\sqrt{(V_{1}+V_{3})^{2}+V_{2}^{2}}/2$, ${\cal
B}=\sqrt{(V_{1}-V_{3})^{2}+V_{2}^{2}}/2$, ${\cal
S}=(V_{1}^{2}+V_{2}^{2}+V_{3}^{2}+V_{4}^{2})/2$, and ${\cal
T}=\sqrt{(V_{1}^{2}+V_{2}^{2}-V_{3}^{2}-V_{4}^{2})^{2}+4\,V_{2}^{2}V_{3}^{2}}/2.$}
\end{table}

\end{document}